\documentclass[preprint,showpacs,preprintnumbers,amsmath,amssymb,prb]{revtex4}

\usepackage{graphicx}% Include figure files
\usepackage{dcolumn}% Align table columns on decimal point
\usepackage{bm}% bold math

\begin{document}

\preprint{Phys. Rev. B}

\title{Polarizational stopping power of heavy-ion diclusters\\ in two-dimensional electron liquids}% Force line breaks with \\

\author{D. Ballester}
 \email{dabalber@mat.upv.es}
\author{A. M. Fuentes}%
\author{I. M. Tkachenko}%
\affiliation{%
Departament de Matem\`{a}tica Aplicada\\ Universitat Polit\`{e}cnica de Val\`{e}ncia, 46022 Valencia, Spain}%

\date{\today}% It is always \today, today,

\begin{abstract}
The in-plane polarizational stopping power of heavy-ion diclusters in a 2D strongly
coupled electron liquid is studied. Analytical expressions for the stopping power of both fast and slow projectiles are derived. To go
beyond the \textit{random-phase approximation} we make use of the inverse
dielectric function obtained by means of the method of moments and some
recent analytical expressions for the static local-field correction factor.
\end{abstract}

\pacs{52.40.Mj, 52.27.Gr, 73.20.Mf}% PACS, the Physics and Astronomy
\maketitle

\section{Introduction}

Investigation of static and dynamic properties of (quasi-)two-dimensional systems of
charged particles has become an important research
area in condensed matter as well as in physics of strongly coupled many-body systems.

Nowadays, it is possible to find a number of examples of experimental realizations of systems
consisting of electrons which are confined within a two-dimensional
configuration. The most relevant example in the classical statistical regime could be that of electrons
trapped in the liquid helium surface.\cite{GA76} But a major number of those experimental realizations correspond to quantal electronic systems \cite{Abrahams} appearing in the fabrication of nanoelectronic devices, such as, e.g., the semiconductor-insulator junctions of semiconductor heterostructures.

On the other hand, the stopping power permits to
characterize the interaction of charged particles with matter and, therefore, it has become a useful diagnostic
tool for experiments with polarizable media. Although,
there are several linear and non-linear physical phenomena contributing to
the loss of energy of moving charged particles interacting with condensed
matter,\cite{Barkas,N95,KP95,BPE99,ZNE2003,ZNE2005,March2005} here we focuse on the part of the energy losses due to the medium polarization. In this physical picture, the loss of energy of the external
projectile is caused by the creation of a dipole between the moving
projectile and the center of the cloud of induced charge which
intends to screen the former.\cite{OT2001} Following the work of Lindhard,\cite{Lindhard}
the calculation of the polarizational stopping power is
usually related to the imaginary part of the medium inverse dielectric function.

In the sequel, we consider a two-dimensional electron gas (2DEG) consisting
of $n$ electrons per unit area immersed in a uniform and rigid neutralizing
background of positive charges. We will assume electrons to interact via a
3D Coulomb potential inversely proportional to the distance.

In the completely degenerate case, such a system can be characterized by a single parameter, which is proportional to the Wigner-Seitz radius, $a$, i.e., the Brueckner parameter%
\begin{equation}
r_{s}=\frac{a}{a_{B}}=\frac{1}{\sqrt{\pi n}a_{B}},  \label{rs}
\end{equation}%
where $a_{B}$ is the Bohr radius.

In addition, this Brueckner parameter can be understood as a coupling parameter, i.e., as a ratio between the characteristic Coulomb interaction energy, $e^{2}/a$, and the characteristic kinetic energy which is the Fermi energy,%
\begin{equation}
E_{F}=\frac{\hbar ^{2}k_{F}^{2}}{2m},  \label{EF}
\end{equation}%
$k_{F}=\left( 2\pi n\right) ^{1/2}$ and $m$ being the Fermi wavenumber and the
electron mass.

In a finite-temperature system, the characteristic kinetic energy is proportional to the temperature in energy units, $\beta^{-1}$, yielding the coupling parameter%
\begin{equation}
\Gamma =\frac{\beta e^{2}}{a}.  \label{Gamma}
\end{equation}%
In this case, an additional parameter is needed to describe the electron fluid, the one which quantifies its degeneracy,%
\begin{equation}
D=\beta E_{F}.  \label{D}
\end{equation}

The strongly coupled regime is usually defined by a value of the coupling
parameter higher than unity. Under such conditions, the mean-field
theories, such as the \textit{random-phase approximation} (RPA), are unable
to account for interparticle correlational effects and, generally speaking, fail to describe the static and dynamic properties of these systems properly. Here the Wigner crystallization \cite{GA79,TC89} constitutes a natural limitation for our approach.

The in-plane polarizational stopping power of a two-dimensional Coulomb
system was considered by Bret and Deutsch, who obtained an expression
relating this quantity, for the case of a general extended
projectile, with the loss function of the system.\cite{BD93} Perhaps, the most outstanding feature of this problem is that the single-ion stopping power in the fast-projectile limiting case exhibits a characteristic asymptotic form which is inversely proportional to the projectile velocity,\cite{BD93} in contrast to the well-known BBL formula applicable in 3D fluids.\cite{BBL} This result has been confirmed by other authors by means of the linear response theory,\cite{WM95} the scattering theory,\cite{N95,KP95,BNE97} and the harmonic oscillator model.\cite{WM96}

In particular, for a dicluster-like charged particle distribution, the expression of the in-plane polarizational stopping power reduces to \cite{BD98}

\begin{equation}
-\frac{dE}{dx}=\left( -\frac{dE}{dx}\right) _{\mathrm{uncorr}}+\left( -\frac{%
dE}{dx}\right) _{\mathrm{corr}}  \label{BD1}
\end{equation}%
\begin{equation}
\left( -\frac{dE}{dx}\right) _{\mathrm{uncorr}}=\frac{2e^{2}}{\pi v}\left(
Z_{1}^{2}+Z_{2}^{2}\right) \int\limits_{0}^{\infty
}dk\int\limits_{0}^{kv}d\omega \frac{\omega }{\sqrt{(kv)^{2}-\omega ^{2}}}%
{\rm Im}\left( -\frac{1}{\varepsilon \left( k,\omega \right) }\right) ,
\label{BD1punct}
\end{equation}%
\begin{equation}
\left( -\frac{dE}{dx}\right) _{\mathrm{corr}}=\frac{2e^{2}}{\pi v}%
2Z_{1}Z_{2}\int\limits_{0}^{\infty }dk\int\limits_{0}^{kv}d\omega \frac{%
\omega J_{0}\left( kR\right) }{\sqrt{(kv)^{2}-\omega ^{2}}}{\rm Im}\left( -%
\frac{1}{\varepsilon \left( k,\omega \right) }\right) ,  \label{BD1corr}
\end{equation}%
where the first term accounts for the stopping power of two point-like
uncorrelated external projectiles with charge numbers $Z_{1}$ and $Z_{2}$,
respectively, whereas (\ref{BD1corr}) stands for the correlated
contribution. We assume here that the two-point-like projectiles are
randomly oriented (see Refs. \onlinecite{BD93ext,BD98}).

In addition, in Ref. \onlinecite{BD93}, the 2D RPA dielectric function was derived and the
asymptotic forms for the polarizational stopping power corresponding to the limiting
 cases of fast and slow single point-like projectiles and diclusters were obtained.\cite{BD98} Later, in the
case of single point-like projectiles,\cite{WM95} an interpolation
formula for the local-field correction factor for a zero-temperature 2DEG
obtained in Ref. \onlinecite{Gold93} was used to account for the electronic correlations.

Here, our main aim is two-fold: first, by means of the method of moments we
show analytically that the fast dicluster stopping power
asymptote of a 2DEG remains unaffected by correlational effects, like it was done
in Ref. \onlinecite{BFT2006} for a single projectile. Second, we revisit the low-velocity asymptote at
zero-temperature, but now including both the single projectile and the
dicluster cases and applying a more recent interpolation formula found
in Ref. \onlinecite{DPGT} which reproduces correctly the short-range effects,\cite{SG88,Holas}
and, finally, we also consider this low-velocity asymptote
in a correlated high-temperature system.\footnote{Some of the results were presented at the 12th International Workshop on the Physics of Non-Ideal Plasmas in Darmstadt, Germany, 2006 (unpublished).}

%%%%%%%%%%%%%%%%%%%%%%%%%%%%%%%%%%%%%%%%%%%%%%%%%%%%%%%%%%%%%%%%%%%%%%%%%%%%%%%%%%%%%%
%%%%%%%%%%%%%%%%%%%%%%%%%%%%%%%%%%%%%%%%%%%%%%%%%%%%%%%%%%%%%%%%%%%%%%%%%%%%%%%%%%%%%%

\section{High-velocity asymptotic forms}

\subsection{Dielectric formalism}

To study the fast-projectile limiting case, we make use of the inverse dielectric
function derived by means of the method of moments and the Nevanlinna formula.\cite{Akh, KN}
The same approach has been applied in the treatment of the
3D and 2D electron gas high-velocity stopping power asymptote,\cite{OT2001,BFT2006} in the case of single-ion projectiles.
In particular, in the 2D case we can write:\cite{OT92}
\begin{equation}
\varepsilon ^{-1}\left( k,z\right) =1+\frac{\omega _{2D}^{2}\left( k\right) %
\left[ z+q\left( k,z\right) \right] }{z\left[ z^{2}-\omega _{2}^{2}\left(
k\right) \right] +q\left( k,z\right) \left[ z^{2}-\omega _{1}^{2}\left(
k\right) \right] }.  \label{nevanlinna}
\end{equation}%
where%
\begin{equation}
\omega _{1}^{2}\left( k\right) =C_{2}\left( k\right) /C_{0}\left( k\right)
=\omega _{2D}^{2}  \left( 1-\varepsilon ^{-1}\left( k,0\right)
\right) ^{-1},  \label{omega1}
\end{equation}%
\begin{equation}
\omega _{2}^{2}\left( k\right) =C_{4}\left( k\right) /C_{2}\left( k\right)
=\omega _{2D}^{2}\left( k\right) \left( 1+K\left( k\right) +L\left( k\right)
\right) ,  \label{omega2}
\end{equation}%

\begin{equation}
C_{\nu }= \int_{-\infty }^{\infty }\omega ^{\nu }%
\mathcal{L} (k,\omega) d\omega ,\quad \nu
=0,2,4, \label{Cnu}
\end{equation}%
being the three non-vanishing and non-diverging frequency power moments (sum-rules) satisfied
by the loss function $\mathcal{L} (k,\omega) = \omega^{-1}{\rm Im}\left( -\varepsilon ^{-1}\left(
k,\omega \right) \right) $:\cite{OT92}

\begin{equation}
C_{0}\left( k\right) = \pi (1-\varepsilon ^{-1}\left( k,0\right)) ,  \label{C0}
\end{equation}%
\begin{equation}
C_{2}\left( k\right) =\pi \frac{2\pi ne^{2}}{m}k\equiv \pi \omega _{2D}^{2},
\label{C2}
\end{equation}%
where $\omega _{2D}=\omega _{2D}\left( k\right) $ is the 2D plasma
frequency, while%
\begin{equation}
C_{4}\left( k\right) =\pi \omega _{2D}^{4}\left( 1+K\left( k\right) +L\left(
k\right) \right) ,  \label{C4}
\end{equation}%
with%
\begin{equation}
K\left( k\right) =\frac{3E_{kin}}{2\pi ne^{2}}k+\frac{\hbar ^{2}}{8\pi
ne^{2}m}k^{3}  \label{K}
\end{equation}%
being the kinetic contribution to the fourth sum-rule ($E_{kin}$ is the
average kinetic energy per electron) and%
\begin{equation}
L\left( k\right) =\frac{1}{N}\sum_{\mathbf{q}(\neq 0,\neq \mathbf{k})}\frac{%
\left( \mathbf{k}\cdot \mathbf{q}\right) ^{2}}{k^{3}q}\left[ S\left( \mathbf{%
k}+\mathbf{q}\right) -S\left( q\right) \right] ,  \label{L}
\end{equation}%
being the correlational contribution.

The parameter function $q\left( k,z\right) $ is an analytic (in the upper
half-plane) Nevanlinna-class function with some specific mathematical properties.\cite{Akh,KN}
From a phenomenological viewpoint the easiest way to choose this
function is to put $q\left( k,z\right) =i0^{+}$. This is equivalent to assume the
existence of a perfectly defined single collective excitation, which acts as
the main energy tranfer channel from the projectile to the plasma. Under
this {\it Ansatz} we should write the loss function as%
\begin{equation}
\mathcal{L} \left( k,\omega \right) =\pi %
\left[ \frac{\omega _{2D}^{2}}{2\omega _{2}^{2}}\delta \left( \omega +\omega
_{2}\right) +\frac{\left( \omega _{2}^{2}-\omega _{1}^{2}\right) C_{0}}{%
\omega _{2}^{2}}\delta \left( \omega \right) +\frac{\omega _{2D}^{2}}{%
2\omega _{2}^{2}}\delta \left( \omega -\omega _{2}\right) \right] ,
\label{canonical}
\end{equation}%
which resembles the Feynman approximation for the dynamic structure
factor.\cite{QLCA,BFT2006} This loss function (\ref{canonical}) not only satisfies all three sum-rules (\ref{C0}-\ref{C4}), but also describes a collective excitation of frequency $\omega_{2}(k)$ which incorporates correlational effects (by means of the contribution (\ref{L})) beyond the RPA.\cite{BFT2006}

\subsection{Stopping power}

Physically, the applicability of the canonical form (\ref{canonical}) of the loss function relies on the fact that for the fast-projectile asymptote the main energy transfer mechanism from the external ion consists in the creation of plasmons in the dielectric medium. This assumption permits to handle this stopping power asymptote analytically even beyond the RPA.

In fact, by applying the loss function (\ref{canonical}), it has
been established that the fast-projectile asymptote of the polarizational
stopping power for point-like projectiles is \cite{BFT2006}

\begin{equation}
\left( -\frac{dE}{dx}\right) _{\mathrm{uncorr}}\simeq \frac{\pi \sqrt{2}%
\left( Z_{1}^{2}+Z_{2}^{2}\right) e^{2}}{r_{s}a_{B}^{2}}\frac{v_{F}}{v},
\label{asymptotic1}
\end{equation}%
which coincides with the result obtained by Bret and Deutsch \cite{BD93,BD98}
in the \textit{random-phase approximation}. Indeed, the arguments used in Ref. \onlinecite{BFT2006} are of a more general nature and are straightforwardly applicable to an external projectile with a general extended distribution either in the 3D case or in the 2D one.\cite{BD93ext,BD93}

%%%%%%%%%%%%%%%%%%%%%%%%%%%%%%%%%%%%%%%%%%%%%%%%%%%%%%%%%%%%
Following the derivation of Ref. \onlinecite{BFT2006}, in the {\it random-phase approximation} we can write the collective excitation frequency appearing in (\ref{canonical}) as

\begin{equation}
\omega _{2}^{2}=\omega _{2D}^{2}\left[ 1+\frac{3E_{kin}^{RPA}}{2\pi ne^{2}}k+\frac{%
\hbar ^{2}}{8\pi ne^{2}m}k^{3}\right] ,  \label{resonance1}
\end{equation} where we neglect correlational contributions in (\ref{omega2},\ref{canonical}). Typically, this approximation is applicable in the weakly 2DEG. 

If we account only for the leading terms of the long- and short-wavelength asymptotes, we get

\begin{equation}
\omega _{2}^{2}\approx \omega _{2D}^{2}\left[ 1+\frac{\hbar ^{2}}{8\pi
ne^{2}m}k^{3}\right] .  \label{resonance2}
\end{equation}

This expression together with (\ref{canonical}) can be introduced into (\ref{BD1corr}) to give

\begin{equation}
\left( -\frac{dE}{dx} \right)_{\rm corr} =\frac{2\sqrt{2} (2 Z_{1} Z_{2}) e^{2}}{r_{s}a_{B}^{2}}\int_{z_{min
}}^{z_{max }}\frac{b^{2}J_{0}\left( 2k_{F} R z\right)}{\sqrt{1-\delta b^{2}z^{-1}-b^{2}z^{2}}}d z,
\label{BD2corr}
\end{equation}%
where $\delta =(2k_{F}a_{B})^{-1}$,  $b\mathfrak{=}v_{F}/v$. In addition, $z_{min}$, $z_{max }$, $z_{min}<z_{max }$, stand for the two positive roots of the polynomial equation $z^{3}-b^{-2}z+\delta =0$,\cite{WM95} which has one negative and two positive roots, $z_{min}$, $z_{max }$, $z_{min}<z_{max }$, if

\begin{equation}
b\ll \delta^{-2/3}. \label{conditionb}
\end{equation}

The asymptotic expansion of the correlated contribution of the dicluster stopping power in 2D (\ref{BD2corr}) gives

\begin{equation}
\left( -\frac{dE}{dx}\right) _{\mathrm{corr}}\simeq \frac{\pi \sqrt{2}\left(
2Z_{1}Z_{2}\right) e^{2}}{r_{s}a_{B}^{2}}J_{0}^{2}\left( \frac{R}{2R_{c}} \right) \frac{v_{F}}{v},  \label{asymptotic2}
\end{equation}%
which coincides with the result obtained in Ref. \onlinecite{BD98} and where

\begin{equation}
R_{c}=\frac{1}{2k_{F} \frac{v}{v_{F}}} = \frac{\hbar}{2mv} \label{Rc}
\end{equation}%
is the fast-projectile coagulation distance.\cite{BD93ext,BD98} In the expansion (\ref{asymptotic2}) we have assumed the ratio $r/b$, with $r=k_{F}R$ and $b=v_{F}/v$ to be finite. But it is clear that when the intercluster distance becomes larger than the Fermi radius $k_{F}^{-1}$, the asymptotic form (\ref{asymptotic2}) is proportional to $v^{-3}$.

In case of a strongly coupled electron fluid, one needs to go beyond the RPA and account for interparticle correlations, i.e., to consider the expression for the collective excitation frequency $\omega_{2}\left( k\right) $ steming from (\ref{omega2}). Here we can again make use of an interpolation formula between the long- and short-wavelength asymptotes, $\omega _{2}^{2}\left( k\downarrow 0\right)$ and $\omega _{2}^{2}\left( k\uparrow \infty \right)$, respectively:\cite{BFT2006}

\begin{equation}
\omega _{2}^{2}\left( k\right) \approx \omega _{2D}^{2}\left[ 1+\frac{%
3E_{kin}}{\pi ne^{2}}k+\frac{5E_{c}}{16\pi ne^{2}}k+\frac{\hbar ^{2}}{8\pi
ne^{2}m}k^{3}\right] ,  \label{omega2_interpol}
\end{equation}%
$E_{c}$ being the correlation energy per particle.

If we introduce expression (\ref{canonical}), but now with (\ref{omega2_interpol}), into (\ref{BD1corr}), then

\begin{equation}
-\frac{dE}{dx}=\frac{2\sqrt{2}(2 Z_{1} Z_{2})e^{2}}{r_{s}a_{B}^{2}}\int_{z_{min
}}^{z_{max }}\frac{b^{2} J_{0}\left( 2k_{F} R z \right)}{\sqrt{1-\xi b^{2}-\delta b^{2}z^{-1}-b^{2}z^{2}}}%
d z,  \label{BD3corr}
\end{equation}%
with
\begin{equation}
\xi =3\frac{E_{kin}}{E_{F}}+\frac{5}{16}\frac{E_{c}}{E_{F}}  \label{xi}
\end{equation}%
which accounts for the contribution of the average kinetic and correlation
energies per electron.

Obviously, this integral can be recast into

\begin{equation}
-\frac{dE}{dx}=\frac{2\sqrt{2}(2 Z_{1} Z_{2})e^{2}}{r_{s}a_{B}^{2}}\sqrt{1-\xi b^{2}}%
\int_{\bar{z}_{min }}^{\bar{z}_{max }}\frac{\bar{b}^{2} J_{0}\left( 2k_{F} R z \right)}{\sqrt{1-\delta
\bar{b}^{2}z^{-1}-\bar{b}^{2}z^{2}}} d z,  \label{BD4corr}
\end{equation}%
with

\begin{equation}
\bar{b}^{2}=\frac{b^{2}}{1-\xi b^{2}},  \label{bbar}
\end{equation} and where condition (\ref{conditionb}) is replaced by

\begin{equation}
b\ll \frac{1}{\sqrt{\xi+\delta^{4/3}}},  \label{conditionbbar}
\end{equation}
%%%%%%%%%%%%%%%%%%%%%%%%%%%%%%%%%%%%%%%%%%%%%%%%%%%%%%%%%%%%
Clearly, the leading term of the asymptotic expansion of (\ref{BD4corr}) coincides with (\ref{asymptotic2}).
Therefore, correlational effects beyond this approach, which are accounted
for in the canonical form (\ref{canonical}) by means of (\ref{L}), do not
affect the asymptotic behavior of the polarizational stopping power obtained
in Refs. \onlinecite{BD93,BD98}.

The expressions (\ref{asymptotic1},\ref{asymptotic2}) have no finite limit when $\hbar \to 0$.\cite{BD93} Notice that the contribution to the dispersion law steming from the electron-hole excitation in (\ref{K}), which is proportional to $k^{3}$, is the main factor responsible for the convergence of the integrals (\ref{BD1punct},\ref{BD1corr}) in the fast projectile limiting case.\cite{BD93,WM95,BFT2006} Then, even for a high-temperature electron liquid, the minimum Coulomb impact parameter which is usually introduced as an upper cut-off in the integration over the wavenumber,%
\begin{equation}
k_{max}^{\prime} \propto \frac{\mu v^{2}}{e^{2}}, \label{kmax1}
\end{equation}%
$\mu$ being the reduced mass, exceeds the de Broglie wavelength and, hence, must be replaced by the latter,\cite{Ichimaru} i.e.,%
\begin{equation}
k_{max} = \frac{2 \mu v}{\hbar}. \label{kmax2}
\end{equation}%
However, the inclusion of the electron-hole excitation term guarantees the convergence of the above-mentioned integrals without introducing the cut-off.

%%%%%%%%%%%%%%%%%%%%%%%%%%%%%%%%%%%%%%%%%%%%%%%
%%%%%%%%%%%%%%%%%%%%%%%%%%%%%%%%%%%%%%%%%%%%%%%

\section{Low-velocity asymptotic forms}

Apparently, the stopping of low-velocity projectiles due to linear
polarizational effects was first studied by Fermi and Teller,\cite{FT47}
who found a characteristic linear dependence of the stopping power on the
projectile velocity. This result seems to be valid for any degree of
degeneracy of the electron fluid and any coupling regime. In fact, this is
just a consequence of the heavy projectile approximation, $M\rightarrow \infty $.\cite{OT2001}
It must be pointed out that, since we are assuming the trajectory
of the projectile to be a straight line, its kinetic energy must always be much greater than the kinetic energy of the
electrons, or, equivalenty,
\begin{equation*}
v>>\sqrt{\frac{m}{M}}v_{F}.
\end{equation*}%
Thus, when we refer to the slow projectile limit we must assume the condition%
\begin{equation*}
\sqrt{\frac{m}{M}}<<\frac{v}{v_{F}}<<1
\end{equation*}%
to hold.\cite{BD93}

\subsection{Dielectric formalism}

In the slow-projectile limiting case, contrary to the previous one, we expect the velocity of the projectile to be too small to create a plasmon in the electron liquid. Thus, now the energy transfer to the plasma is essentially determined by the low-frequency range of the fluctuation spectrum. In addition, in spite of the efforts to find an adequate expression for the parameter function $q(k,\omega)$ appearing in (\ref{nevanlinna}), capable to describe this range of the spectrum \cite{BT2005} adequately, here we make use of the general static local-field-corrected dielectric function,%
\begin{equation}
\varepsilon \left( k,\omega \right) =1+\phi \left( k\right) \Pi \left(
k,\omega \right) =1+\frac{\phi \left( k\right) \Pi _{RPA}\left( k,\omega
\right) }{1-\phi \left( k\right) G\left( k\right) \Pi _{RPA}\left( k,\omega
\right) },  \label{epsilon_gen}
\end{equation}%
which permits to apply recent numerical results on the static properties of the electron gas obtained by numerical simulations. In the previous expression $\phi \left( k\right) =2\pi e^{2}/k$ is the two-dimensional Coulomb
potential, $\Pi \left( k,\omega \right) $ is the polarization function of
the system, $\Pi _{RPA}\left( k,\omega \right) $ its form within the RPA,
and $G\left( k\right) $ is the static local-field correction function.

Further, from expression (\ref{epsilon_gen}) we can derive the loss function:

\begin{equation}
\mathcal{L} \left( k,\omega \right) =\frac{1}{%
\omega }\frac{\phi \left( k\right) \Pi _{RPA}^{\prime \prime }\left(
k,\omega \right) }{\left[ 1+\phi \left( k\right) H\left( k\right) \Pi
_{RPA}^{\prime }\left( k,\omega \right) \right] ^{2}+\left[ \phi \left(
k\right) H\left( k\right) \Pi _{RPA}^{\prime \prime }\left( k,\omega \right) %
\right] ^{2}},  \label{lossfunc}
\end{equation}%
where we define the function $H\left( k\right) =1-G\left( k\right) $
and $\Pi _{RPA}\left( k,\omega \right) =\Pi _{RPA}^{\prime }\left( k,\omega
\right) +i\Pi _{RPA}^{\prime \prime }\left( k,\omega \right) $. Obviously,
to recover the RPA we must put $H\left( k\right) =1$.

\subsection{Stopping power}

Next, we introduce this last expression (\ref{lossfunc}) into the general equations (\ref{BD1punct})
and (\ref{BD1corr}). In order to simplify the corresponding
integrals, we can make use of two common approximations: (i) since we are
dealing with slow projectiles, $v/v_{F}\ll 1$, we substitute the denominator of the loss function by its static
limiting form ($\omega =0$) and recall that $\Pi
_{RPA}^{\prime \prime }\left( k,0\right) =0$; (ii) we can approximate in
the numerator the imaginary part of the polarization function $\Pi
_{RPA}^{\prime \prime }\left( k,\omega \right) $ as$\ \omega \left( \partial
\Pi _{RPA}^{\prime \prime }/\partial \omega \right) _{\omega =0}$.

\subsubsection{Zero-temperature system}

In the zero-temperature limiting case, the slow-projectile expression corresponding
to the point-like contribution (\ref{BD1punct}), can be recast as

\begin{equation}
\left( -\frac{dE}{dx}\right) _{\mathrm{uncorr}}=\frac{2\sqrt{2}\left(
Z_{1}^{2}+Z_{2}^{2}\right) e^{2}}{r_{s}a_{B}^{2}}\frac{v}{v_{F}}%
\int\limits_{0}^{1}\frac{dz}{\sqrt{1-z^{2}}}\frac{z^{2}}{\left[ z+\frac{r_{s}%
}{\sqrt{2}}H\left( z\right) \right] ^{2}}.  \label{stopp_slow2_punct}
\end{equation}%
which corresponds to that derived in Ref. \onlinecite{WM95} for a point-like
projectile, whereas for the contribution (\ref{BD1corr}) we have%
\begin{equation}
\left( -\frac{dE}{dx}\right) _{\mathrm{corr}}=\frac{2\sqrt{2}\left(
2Z_{1}Z_{2}\right) e^{2}}{r_{s}a_{B}^{2}}\frac{v}{v_{F}}\int\limits_{0}^{1}%
\frac{dz}{\sqrt{1-z^{2}}}\frac{z^{2}J_{0}\left( 2k_{F}Rz\right) }{\left[ z+%
\frac{r_{s}}{\sqrt{2}}H\left( z\right) \right] ^{2}}.
\label{stopp_slow2_corr}
\end{equation}%
These two expressions generalize those derived by Bret and Deutsch for slow
projectiles.\cite{BD98} Again, these are recovered under the RPA, i.e.,
with $H\left( k\right) =1$.

On the other hand, non-linear (non-perturbative) stopping power effects beyond our (linear) polarizational approach might be important in the slow-projectile limiting case. These effects have been treated using the non-perturbative scattering theory in a 
partial-wave representation, both for single heavy ions \cite{N95,KP95,ZNE2003,ZNE2005} and diclusters.\cite{March2005} In particular, for randomly oriented slow diclusters of similar 'atoms', in Ref. \onlinecite{March2005} it was shown that the correlated contribution to the differential cross section, appears via a Bessel 
function factor similar to that of (\ref{stopp_slow2_corr}).

To go beyond the RPA we need an analytical expression for the
local-field correction factor, $G\left( k\right) $, of a paramagnetic
two-dimensional degenerate electron gas. For instance, in Ref. \onlinecite{WM95}
an interpolation formula obtained in Ref. \onlinecite{Gold93} was used. In additon, a
more recent interpolation formula can be found in Ref. \onlinecite{DPGT}. Indeed,
this latter expression not only accounts for the well-known long-wavelength
asymptotic behavior of the local-field correction factor, but also
reproduces correctly the short-range effects.\cite{SG88,Holas} In
particular,%
\begin{equation}
G\left( k\downarrow 0\right) \simeq A\frac{k}{k_{F}},  \label{Glong}
\end{equation}%
and the constant $A$ is determined via the compressibility sum-rule to
give \cite{DPGT}%
\begin{equation}
A=\frac{1}{\sqrt{2}r_{s}}\left( 1-\frac{\kappa _{0}}{\kappa }\right) ,
\label{A}
\end{equation}%
where $\kappa _{0}=\pi r_{s}^{4}/2$\ is the compressibility of the ideal
electron gas (in the units of $a_{B}^{2}/Ry$), whereas $\kappa $ is the
compressibility of the interacting gas. For $T=0$, we have%
\begin{equation}
\frac{\kappa _{0}}{\kappa }=1-\frac{\sqrt{2}}{\pi }r_{s}+\frac{r_{s}^{4}}{8}%
\left( \frac{d^{2}E_{c}\left( r_{s}\right) }{dr_{s}{}^{2}}-\frac{1}{r_{s}}%
\frac{dE_{c}\left( r_{s}\right) }{dr_{s}{}}\right) .  \label{kappa}
\end{equation}%
Here, for the correlation energy per particle, $E_{c}$, one might use, for
instance, the expression obtained by Rapisarda and Senatore \cite{RS96} by
fitting of data of the diffusion-Monte-Carlo simulations or the one fitted
from the quantum-Monte-Carlo simulation results in Ref. \onlinecite{AMGB}:%
\begin{eqnarray}
\frac{a_{B}E_{c}\left( r_{s}\right) }{e^{2}}
&=&-0.1925+(0.0863136r_{s}+0.0572384r_{s}^{2}+0.00363r_{s}^{3})
\label{correlationenergy} \\
&&\times \ln \left[ 1+\left(
1.0022r_{s}-0.02069r_{s}^{3/2}+0.33997r_{s}^{2}+0.01747r_{s}^{3}\right) ^{-1}%
\right] ,  \notag
\end{eqnarray}%
which has been obtained for $1\leq r_{s}\leq 40$ (see also Ref. \onlinecite{APDT}
and references therein for a further discussion).

On the other hand, for the short-wavelength asymptotic form we have \cite{SG88,Holas}%
\begin{equation}
G\left( k\uparrow \infty \right) \simeq C\frac{k}{k_{F}}+B,  \label{Gshort}
\end{equation}%
where $C$ depends linearly on the difference in kinetic energy between the
interacting and the ideal electron gas,\cite{DPGT}%
\begin{equation}
C=-\frac{r_{s}}{2\sqrt{2}}\frac{d}{dr_{s}{}}\left( r_{s}E_{c}\left(
r_{s}\right) \right) .  \label{C}
\end{equation}

In addition, $B=1-g\left( 0\right) $, where $g\left( 0\right) $\ is the
value of the pair-correlation function at the origin. For the latter we can make use
of the interpolation formula \cite{PSDT}%
\begin{equation}
g\left( 0\right) =\frac{1/2}{1+1.372r_{s}+0.0830r_{s}^{2}}.  \label{g0}
\end{equation}

\begin{figure*}
\centering{
\includegraphics[width=.28\textwidth]{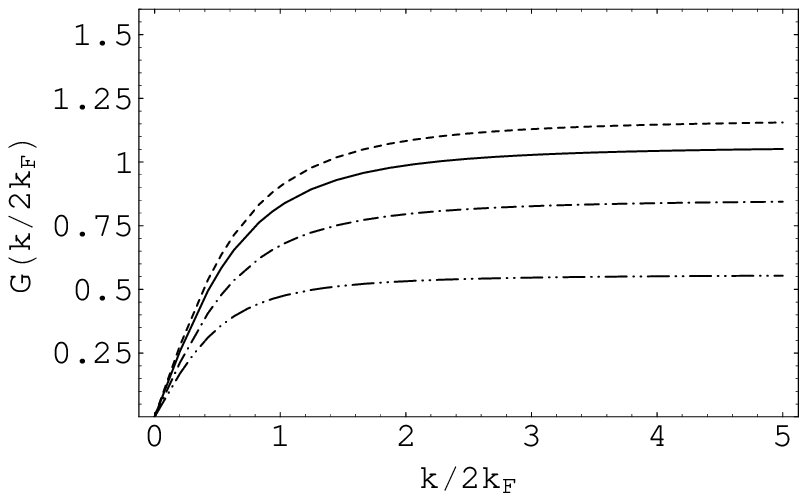}~(a)
\hfil
\includegraphics[width=.28\textwidth]{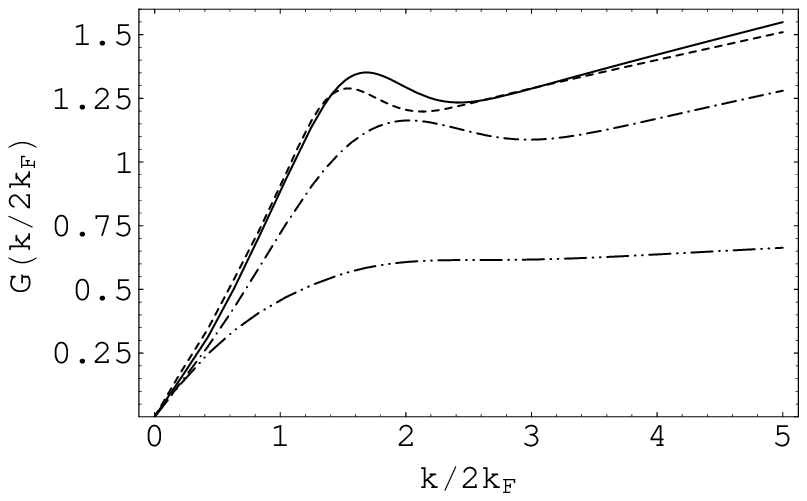}~(b)
\hfil
\includegraphics[width=.28\textwidth]{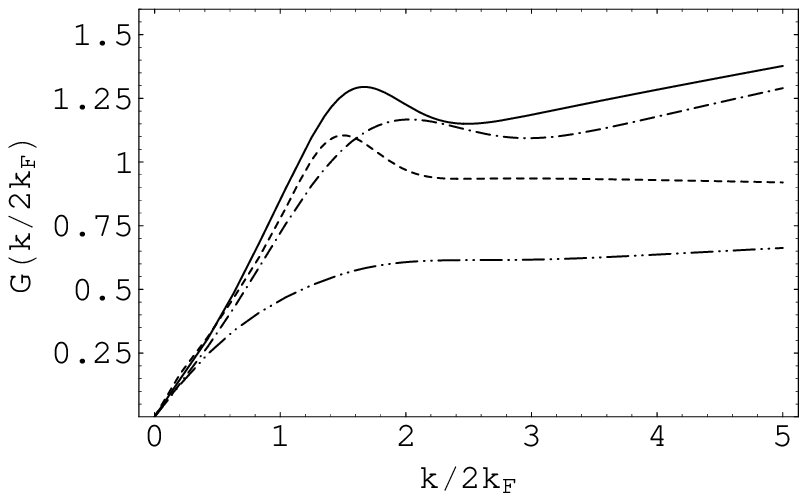}~(c)}
\caption{}
\label{fig:gT0}
\end{figure*}

Although the interpolation formula obtained in Ref. \onlinecite{DPGT} for the
static local-field correction factor was explictly derived by using the correlation
energy formula of Ref. \onlinecite{RS96} valid in the range $0.1\leq r_{s}\leq 10$, here 
we also use this expression with the interpolation formula (\ref{correlationenergy}) for the correlation energy
obtained in Ref. \onlinecite{AMGB}. The results for the
static local-field correction are compared in Fig. \ref{fig:gT0}. Notice that the expressions obtained by means of the interpolation formula of Ref. \onlinecite{DPGT} show the shift of the peaks beyond $2k_{F}$, as decribed in Ref. \onlinecite{DHW}.

\begin{figure}
\centering{
\includegraphics{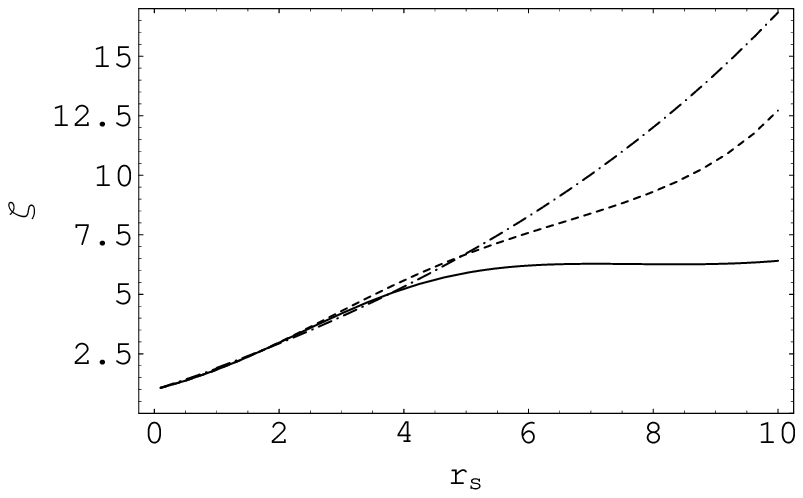}}
\caption{}
\label{figdzeta}
\end{figure}

In Fig. \ref{figdzeta} we plot the ratio between the low-velocity asymptotic form of the stopping power of a single projectile, i.e., the sum of (\ref{stopp_slow2_punct}) and (\ref{stopp_slow2_corr}) with $R=0$, and its RPA counterpart,

\begin{equation}
\zeta =\left( -\frac{dE}{dx}\right) \left( -\frac{dE}{dx}\right)
_{RPA}^{-1}\geq 1,  \label{dzeta}
\end{equation}%
and study the difference obtained using the interpolation formula of Ref. \onlinecite{DPGT}
(with the expression for the correlation energy of Ref. \onlinecite{RS96} or of Ref. \onlinecite{AMGB}, respectively) or that of Ref. \onlinecite{Gold93}, as it was done in Ref. \onlinecite{WM95}.

As one can see, the electronic correlations enhance the stopping power, this increase becoming more pronounced for lower
electron densities (high coupling region), as it was outlined previously.\cite{WM95}

Moreover, although we only need to integrate in (\ref{stopp_slow2_punct},\ref{stopp_slow2_corr}) until the value of $k=2k_{F}$, we see that there is a
certain quantitative discrepancy due to the usage of the expressions for the
correlation energy $E_{c}\left( r_{s}\right) $ from Ref. \onlinecite{RS96} or from Ref. \onlinecite{AMGB}. In both cases, it is also noticeable that the denominator in the
integrand of (\ref{stopp_slow2_punct},\ref{stopp_slow2_corr}) vanishes for the
values of $r_{s}>10$, which is indeed beyond the fitting range of $G\left(
k\right) $ in Refs. \onlinecite{Gold93,APDT}.

It is clear from (\ref{stopp_slow2_corr}) that in the slow-projectile limiting case the natural scale for the distance between the ions in the dicluster is the coagulation distance in a completely degenerate system,\cite{BD93ext,BD98}%
\begin{equation}
R_{c}^{\prime}=\frac{\lambdabar_{F}}{2}=\frac{1}{2 k_{F}}= \frac{a_{B}}{2\sqrt{2}} r_{s}. \label{Rcslow1}
\end{equation}%

\begin{figure*}
\centering{
\includegraphics[width=.40\textwidth]{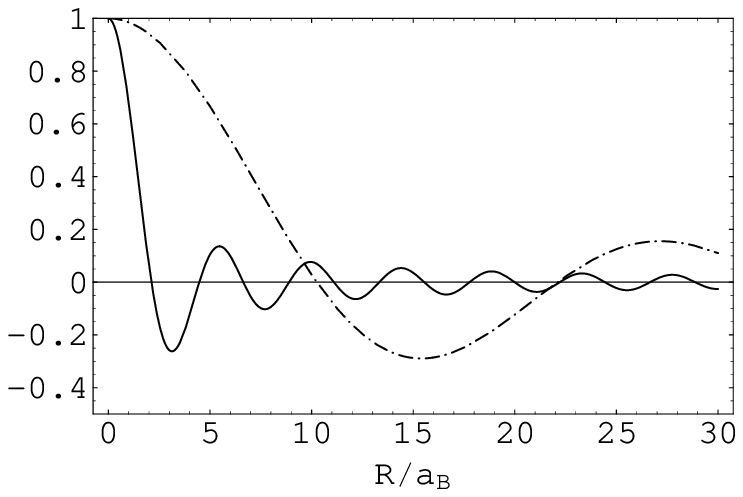}~(a)
\hfil
\includegraphics[width=.40\textwidth]{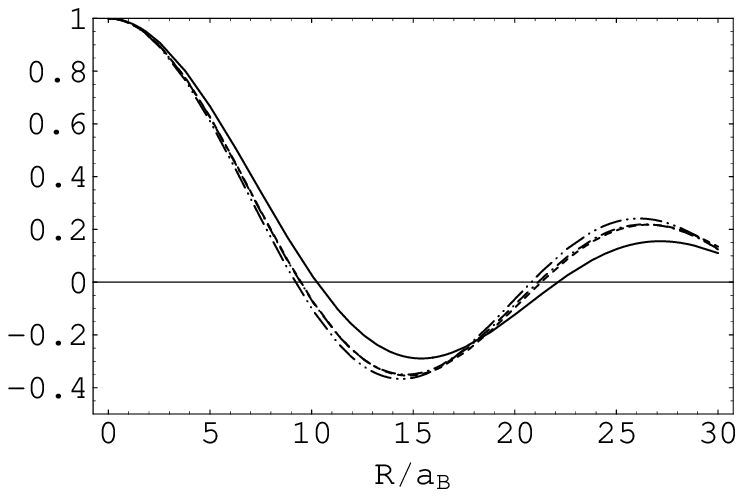}~(b)
}
\caption{}
\label{fig:corrpunctT0}
\end{figure*}

In Fig. \ref{fig:corrpunctT0} we compare the ratio of the correlated low-velocity asymptotic form (\ref{stopp_slow2_corr}) to the uncorrelated one (\ref{stopp_slow2_punct}) of a dicluster distribution as a function of the distance between the ions in atomic units, for different values of the Brueckner parameter, in both the RPA and the beyond-the-RPA cases, and assuming that both ions have the same charge number.

In particular, an increase of the value of the Brueckner parameter makes the Friedel-like oscillatory pattern of (\ref{stopp_slow2_corr}) to exhibit a lower damping rate (Fig. \ref{fig:corrpunctT0} (a)) due to the proportionalty between the Brueckner parameter and the coagulation distance $R_{c}^{\prime}$ (\ref{Rcslow1}).

In addition, the comparison of the above-mentioned ratio of the correlated low-velocity asymptotic form (\ref{stopp_slow2_corr}) to the uncorrelated one (\ref{stopp_slow2_punct}) in the RPA and beyond the RPA, might provide a quantitative insight into the dependence of the screening distance on the value of the Brueckner parameter (Fig. \ref{fig:corrpunctT0} (b)). 

The 2D Thomas-Fermi wavelength, $\lambdabar_{TF}=a_{B}/2$,\cite{BD93} represents a typical screening distance in the {\it random-phase approximation}. In the region where the RPA is valid, $r_{s} \ll 1$, the Wigner-Seitz radius is smaller than $a_{B}$. This means that the total number of particles which are contained inside the disk of radius $\lambdabar_{TF}$,%
\begin{equation}
N=\pi \lambdabar_{TF}^{2}n=\frac{1}{4 r_{s}^{2}}, \label{N1}
\end{equation}
is large enough to screen the potential created by the external ion effectively. On the other hand, beyond the domain of applicability of the RPA, $r_{s}>1$, the number of particles inside the disk of radius $\lambdabar_{TF}$ is smaller than unity and, then, the Thomas-Fermi wavelength becomes an underestimate of the typical screening distance. Thus, we observe that the oscillations exhibited at $r_{s}=10$ in the RPA case in Fig. \ref{fig:corrpunctT0} (b) become even less damped when the local-field correction factor $G(k)$, which permits to describe the behavior of the system at higher values of the Brueckner parameter more accurately, is accounted for.

\subsubsection{High-temperature system}

In the high-temperature limiting case, expression (\ref{BD1punct}) with (\ref{lossfunc}) reduces to

\begin{equation}
\left( -\frac{dE}{dx}\right) _{\mathrm{uncorr}}= \sqrt{\pi}
\left( Z_{1}^{2}+Z_{2}^{2}\right) p v k_{D} \int\limits_{0}^{\infty }dx%
\frac{x^{2}\exp \left( -x^{2}\right) }{\left[ x+\frac{\Gamma \sqrt{D}}{\sqrt{%
2}}H\left( \frac{x}{\sqrt{D}}\right) \right] ^{2}},
\label{stopp_slow3_punct}
\end{equation}%
where $k_{D}=2\pi ne^{2}\beta $ is the 2D Debye wavenumber, $p=me^{2}/\hbar$, whereas (\ref%
{BD1corr}) can be written as

\begin{equation}
\left( -\frac{dE}{dx}\right) _{\mathrm{corr}}=2 \sqrt{\pi} 
Z_{1}Z_{2} pv k_{D} \int\limits_{0}^{\infty }dx\frac{%
x^{2}\exp \left( -x^{2}\right) J_{0}\left( \bar{R}x\right) }{\left[ x+\frac{%
\Gamma \sqrt{D}}{\sqrt{2}}H\left( \frac{x}{\sqrt{D}}\right) \right] ^{2}},
\label{stopp_slow3_corr}
\end{equation}%
with $\bar{R}=2R/\lambdabar$, $\lambdabar=(\hbar^{2}\beta/2m)^{1/2}$ being the thermal de Broglie wavelength. Again, if we put $H\left(
k\right) =1$ we recover the result obtained by Bret and Deutsch for the
limiting form of the stopping power of slow projectiles \cite{BD93,BD98} of
a high-temperature system.

Once more, Eqs. (\ref{stopp_slow3_punct},\ref{stopp_slow3_corr}) have no finite limit as $\hbar \to 0$, although now the previous cut--off argument is not applicable. Nevertheless, one migth argue that to avoid the Coulomb collapse on a slow projectile, one has to take into account the effects of electronic diffraction essentially related to the finiteness of $\hbar$. Notice that this peculiarity, in both the fast-projectile limiting case and the slow-projectile one, is shared by the 3D problem as well,\cite{OT2001} and, therefore, seems to be specific for the calculation of the polarizational contribution to the stopping power.

Since the expressions for the local-field correction factor employed before are only
applicable to zero-temperature systems, to go beyond the RPA we
consider here an interpolation formula based on the known
asymptotes that $G\left( k\right) $ must fulfill.\cite{OT2001}

First, the long-wavelength behavior is described by expression (\ref%
{Glong}), being the value of the constant $A$ again dictated by the
compressibility sum-rule. For a high-temperature system, $D\ll 1$, we have
\cite{OT2001}%
\begin{equation}
A=-\frac{\beta E_{c}\left( \Gamma \right) +\frac{\Gamma }{3}\frac{d}{d\Gamma
}\beta E_{c}\left( \Gamma \right) }{\left( \frac{12}{\pi }\right)
^{2/3}\Gamma },  \label{A_classical}
\end{equation}%
where for the correlation energy per electron of a classical electron gas we
can use the MC formula of Totsuji \cite{Totsuji}%
\begin{equation}
\beta E_{c}\left( \Gamma \right) =-1.12\Gamma +0.71\Gamma ^{1/4}-0.38,
\label{Ec_classical}
\end{equation}%
which is valid for $\sqrt{2}<\Gamma <50$.

\begin{figure}
\centering{
\includegraphics{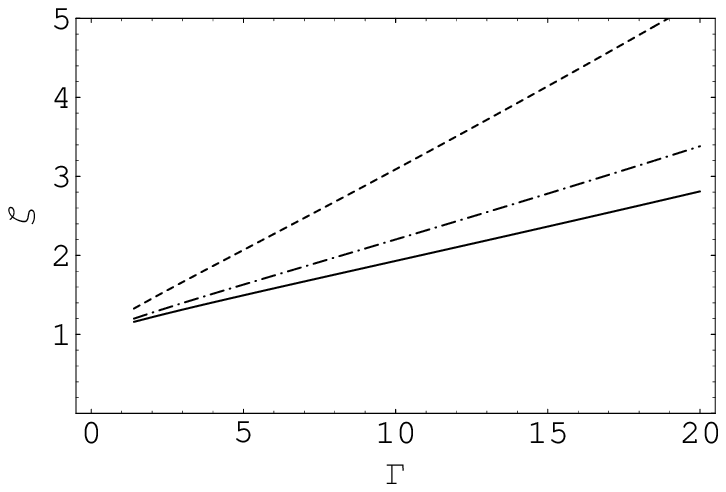}}
\caption{}
\label{figdzeta_classical}
\end{figure}

\begin{figure*}
\centering{
\includegraphics[width=.40\textwidth]{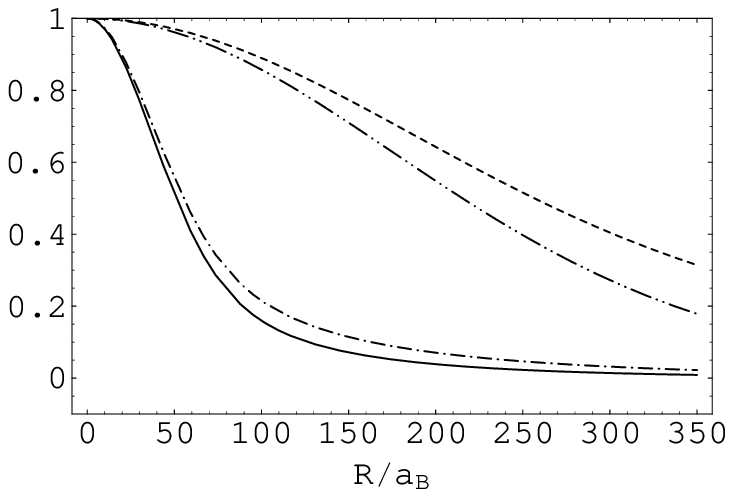}~(a)
\hfil
\includegraphics[width=.40\textwidth]{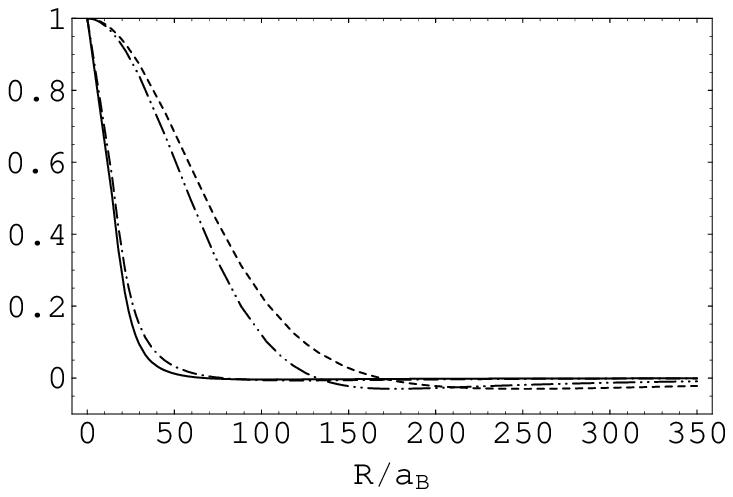}~(b)
}
\caption{}
\label{fig:corrpunctT}
\end{figure*}

For the long-wavelength behavior of the local-field correction
factor in the high-temperature system, one expects the Holas contribution
\cite{Holas} to vanish. Indeed, in the classical limiting case the kinetic energy
of an interacting electron system coincides with that of a non-interacting one.
Hence, we can write%
\begin{equation}
G\left( k\uparrow \infty \right) \simeq 1-g\left( 0\right) ,
\label{Gshort_classical}
\end{equation}%
where we can even assume that the electron-electron pair-correlation function
vanishes at the origin in the high-temperature limiting case. With this in
mind, the interpolated expression for the local-field correction factor can be cast as%
\begin{equation}
G\left( z\right) =\frac{2Az}{1+2Az},  \label{Gclassical}
\end{equation}%
where $z=k/2k_{F}$. This high-temperature static local-field correction is used to plot Figs. \ref{figdzeta_classical} and \ref{fig:corrpunctT} for a dicluster projectile with $Z_{1}=Z_{2}$.

In Fig. \ref{figdzeta_classical} we display the ratio (\ref{dzeta}) for a
high-temperature strongly coupled 2D system at different values of the
degeneracy parameter $D$. As in the completely degenerate case before, one can see that $\zeta
\left( \Gamma ;D\right) \geq 1$, and the departure from the RPA becomes more
notorious as the coupling parameter increases, although the former is slightly smaller at lower values of the degeneracy parameter than the one found in the zero-temperature system, i.e., there is a consistent tendency of enlargement of the ratio $\zeta
\left( \Gamma ;D\right)$ as the degeneracy increases.

In the classical limiting case, $D\ll 1$, the natural length to measure the distance between the cluster ions (see Eq. (\ref{stopp_slow2_corr})) is substituted by the thermal de Broglie wavelength which is proportional to the coagulation distance in the slow-projectile limiting case,\cite{BD93ext,BD98}%
\begin{equation}
R_{c}^{\prime\prime}=\frac{\lambdabar}{2}=\frac{a_{B}}{2} \frac{\Gamma}{\sqrt{2D}}, \label{Rcslow2}
\end{equation}%
since it represents the minimum distance which might be resolved in the electron liquid due to the uncertainty principle. Here the typical screening distance in a weakly coupled plasma, e.g., the Debye radius, is also much larger than the thermal de Broglie wavelength.

Further, due to the relationship expressed in (\ref{Rcslow2}), the behavior of the correlational term (\ref{stopp_slow3_corr}) is similar to that of the zero-temperature system, i.e., its damping rate decreases as the value of the coupling parameter increases, although this tendency appears to be obviously compensated at higher values of the degeneracy (Fig. \ref{fig:corrpunctT}). It is also noticeable that when the degeneracy rises the contribution (\ref{stopp_slow3_corr}) becomes negative and there is a tendency to reproduce the Friedel-like oscillations which characterize the completely degenerate system.

As in the previous case, the evaluation of the dependency of the screening distance on the value of the coupling in the RPA and with a local-field-corrected dielectric function, clearly indicates that the quantitative difference found previously remains applicable at high temperatures. Indeed, under these conditions the number of particles inside the disk of radius $\lambdabar_{D}$,%
\begin{equation}
N=\pi \lambdabar_{D}^{2}n=\frac{1}{4 \Gamma^{2}}, \label{N2}
\end{equation}
is insufficient to screen the disturbance induced by the ion potential, and the typical Debye radius becomes again an underestimate of the screening distance as the coupling increases.

%%%%%%%%%%%%%%%%%%%%%%%%%%%%%%%%%%%%%%%%%%%%%%%%%%%%%%%
%%%%%%%%%%%%%%%%%%%%%%%%%%%%%%%%%%%%%%%%%%%%%%%%%%%%%%%

\section{Conclusions}

Correlational contributions to the in-plane polarizational stopping power of heavy-ion diclusters by 2D strongly coupled electron fluids have been
assessed. The limiting forms for fast- and slow-projectiles and for both
low- and high-temperature cases have been considered. In the case of a
high-velocity projectile we have used a dielectric formalism based on the
employment of the canonical solution of the truncated problem of moments for
the loss function, and have established that the fast-dicluster
stopping power asymptote is unaffected by the interparticle correlations. On
the other hand, for slow projectiles we have compared several fitted
expressions for the static local-field factor. In particular, an \textit{ad-hoc} interpolated formula has been constructed for the classical system.
In this slow-projectile asymptote the correlational effects are shown to be
quantitatively important, as it was previously outlined.

%%%%%%%%%%%%%%%%%%%%%%%%%%%%%%%%%%%%%%%%%%%%%%%%%%%%%%%
%%%%%%%%%%%%%%%%%%%%%%%%%%%%%%%%%%%%%%%%%%%%%%%%%%%%%%%

\begin{acknowledgments}
The authors are grateful to G. Senatore for valuable comments and acknowledge the financial support provided in the framework of the GSI-INTAS Project No. 03-54-4254.
\end{acknowledgments}

%\appendix

% \section{}
% \label{}

%%%%%%%%%%%%%%%%%%%%%%%%%%%%%%%%%%%%%%%%%%%%%%%%%%%%
%%%%%%%%%%%%%%%%%%%%%%%%%%%%%%%%%%%%%%%%%%%%%%%%%%%%

% Bibliographic references with the natbib package:
% Parenthetical: \citep{Bai92} produces (Bailyn 1992).
% Textual: \citet{Bai95} produces Bailyn et al. (1995).
% An affix and part of a reference:
%   \citep[e.g.][Ch. 2]{Bar76}
%   produces (e.g. Barnes et al. 1976, Ch. 2).

\newpage

{\bf Figure Captions}\newline

Fig. 1. The fitted expression for the
static local-field correction factor of: (a) Ref. \onlinecite{Gold93}; (b) Ref. \onlinecite{APDT} with $E_{c}\left(
r_{s}\right) $ from Ref. \onlinecite{RS96}; (c) Ref. \onlinecite{APDT} with $E_{c}\left(
r_{s}\right) $ from Ref. \onlinecite{AMGB}. Dash-dot-dot:
$r_{s}=0.1$; dash-dot: $r_{s}=1$; solid: $r_{s}=5$; dashed: $r_{s}=10$.\newline

Fig. 2. Ratio between the
low-velocity asymptotic expression for the stopping power and its RPA
counterpart (\ref{dzeta}) in the zero-temperature limit and for a
single projectile, i.e., $R=0$. The fitted expression for the static
local-field correction factor of Ref. \onlinecite{APDT} with $E_{c}\left(
r_{s}\right) $ from Ref. \onlinecite{RS96} (dashed) and from Ref. \onlinecite{AMGB}
(solid) are compared to the expression of Ref. \onlinecite{Gold93}
(dash-dot).\newline

Fig 3. Ratio between the correlated low-velocity asymptotic expression (\ref{stopp_slow2_corr}) for
the dicluster stopping power and its uncorrelated counterpart (\ref{stopp_slow2_punct})
in the zero-temperature limiting case with $Z_{1}=Z_{2}$: (a) RPA for $r_{s}=2$ (solid) and $r_{s}=10$ (dash-dot); (b) Comparison of different fitted expressions for the
static local-field correction factor for $r_{s}=10$: RPA, i.e., $G(k)=0$, (solid), $G(k)$ of Ref. \onlinecite{APDT} with $E_{c}\left(
r_{s}\right) $ from Ref. \onlinecite{RS96} (dash-dot-dot), $G(k)$ of Ref. \onlinecite{APDT} with $E_{c}\left(
r_{s}\right) $ from Ref. \onlinecite{AMGB} (dash-dot), and $G(k)$ of Ref. \onlinecite{Gold93} (dashed).\newline

Fig. 4. Ratio between the correlated low-velocity asymptotic expression for
the stopping power and its RPA counterpart (\ref{dzeta}) in the
high-temperature limiting cases and for a single projectile, i.e., $R=0$. The
interpolated expression for the static local-field correction factor (\ref{Gclassical}) is used. The values of the degeneracy parameter, $D$, are: $10^{-3}$ (solid), $2\times 10^{-3}$ (dash-dot), and $10^{-2}$
(dashed).\newline

Fig. 5. Ratio between the correlated low-velocity asymptotic expression (\ref{stopp_slow3_corr}) for
the dicluster stopping power and its uncorrelated counterpart (\ref{stopp_slow3_punct})
in the high-temperature limiting case with $Z_{1}=Z_{2}$ for $D=10^{-3}$ (a) and $D=10^{-2}$ (b). The values
of the coupling parameter are: $\Gamma=2$ (with $G(z)=0$, (solid), or $G(z)$ from (\ref{Gclassical}), (dash-dot)) and $\Gamma=10$ (with $G(z)=0$, (dash-dot-dot), or $G(z)$ from (\ref{Gclassical}), (dashed)).\newline

\newpage

\vspace*{2cm}

\vspace*{2cm}

\vspace*{2cm}

\vspace*{2cm}

\end{document}